# Generative Models Improve Radiomics Reproducibility in Low Dose CTs: A Simulation Study

**Junhua Chen[1], Chong Zhang [1], Alberto Traverso[1], Ivan Zhovannik[1,2], Andre Dekker[1], Leonard Wee[1,3] and Inigo Bermejo[1,3]**

[1] Department of Radiation Oncology (MAASTRO), GROW School for Oncology and Developmental Biology, Maastricht University Medical Centre+, Maastricht, 6229 ET, Netherlands
[2] Department of Radiation Oncology, Radboud Institute for Health Sciences, Radboud University Medical Center, Nijmegen, 6525 GA, Netherlands
[3] joint senior authors

E-mail: Junhua.chen@maastro.nl

## Abstract

Radiomics is an active area of research in medical image analysis, however poor reproducibility of radiomics has hampered its application in clinical practice. This issue is especially prominent when radiomic features are calculated from noisy images, such as low dose computed tomography (CT) scans. In this article, we investigate the possibility of improving the reproducibility of radiomic features calculated on noisy CTs by using generative models for denoising. Our work concerns two types of generative models – encoder-decoder network (EDN) and conditional generative adversarial network (CGAN). We then compared their performance against a more traditional "non-local means" denoising algorithm. We added noise to sinograms of full dose CTs to mimic low dose CTs with two levels of noise: low-noise CT and high-noise CT. Models were trained on high-noise CTs and used to denoise low-noise CTs without re-training. We tested the performance of our model in real data, using a dataset of same-day repeated low dose CTs in order to assess the reproducibility of radiomic features in denoised images. EDN and the CGAN achieved similar improvements on the concordance correlation coefficients (CCC) of radiomic features for low-noise images from 0.87 [95%CI, (0.833,0.901)] to 0.92[95%CI, (0.909,0.935)] and for high-noise images from 0.68 [95%CI, (0.617,0.745)] to 0.92[95%CI, (0.909,0.936)], respectively. The EDN and the CGAN improved the test-retest reliability of radiomic features (mean CCC increased from 0.89 [95%CI, (0.881,0.914)] to 0.94[95%CI, (0.927,0.951)]) based on real low dose CTs. These results show that denoising using EDN and CGANs could be used to improve the reproducibility of radiomic features calculated from noisy CTs. Moreover, images at different noise levels can be denoised to improve the reproducibility using the above models without need for re-training, provided the noise intensity is not excessively greater that of the high-noise CTs. To the authors' knowledge, this is the first effort to improve the reproducibility of radiomic features calculated on low dose CT scans by applying generative models.

## 1. Introduction

Radiomics is currently an active area of research in medical image analysis. It involves the automated extraction of (either hand-crafted or deep-learning) quantitative image metrics known as "features", in the hope of improving the diagnostic, prognostic, or predictive accuracy of clinical models [1].One of the major advantages of hand-crafted radiomics features, as opposed to deep features, is higher potential for interpretability by human operators.

Radiomics has shown potential for clinical-decision support in oncology for a diverse range of cancer sites such as lung [2], head and neck [3], and rectal cancer [4], among others. The most widely used clinical imaging modalities for radiomics are computed tomography (CT) [3][5], magnetic resonance imaging (MRI) [6], and positron emission tomography (PET) [7]. Radiomics has attracted increased attention from researchers following the seminal article by Aerts *et al.* [7] in 2014. In spite of significant progress made during recent years, there remain barriers that hamper widespread adoption of radiomics in clinical settings. One important issue is the generally poor repeatability and reproducibility of radiomic features [8]. Repeatability refers to features that remain the same when extracted multiple times in the same subject. The repeatability of radiomic features may be assessed by test-retest imaging [9] and interobserver studies of tumor contouring [10]. Radiomics reproducibility is the stability of features when at least one processing condition (e.g. equipment, software, acquisition settings) has been changed. The reproducibility of radiomics features is key to external validity and widespread generalizability with respect to differences in image reconstruction [11], radiation dose during CT scanning [12], and other variations that inevitably arise across clinics and scanners.

An important source of non-reproducibility or limited generalizability in radiomics that demands more attention is image noise. Due to the long term risk posed by low levels of ionizing radiation exposure, low dose CTs have become more popular (e.g. the ALARA principle [17]) especially for screening and monitoring of populations at risk. Therefore, researchers have interest to use radiomic features based on low dose CTs. As an inevitable trade-off for low radiation exposure, higher noise is present in these images. This noise decreases the image texture [13]. As reported in [11], changes in radiation dose reduce the reproducibility of radiomic features, and features from low dose CT images tend to have lower reproducibility [12]. Image noise has been shown to adversely impact the reproducibility of radiomic features if signal to noise ratio (SNR) falls below 50 dB [5], but the results also show that some radiomics features are robust to low-pass filtering. Thus, improving the reproducibility of radiomics from low dose CTs is a timely and potentially impactful clinical research topic.

To the best of the authors' knowledge, there are presently no published detailed analysis on how to improve radiomics features robustness in low dose CT. A potential solution worth exploring is pre-extraction denoising of images [12]. Jiang, *et al.* [14] described a new semi-sophisticated generative adversarial network (GAN) that reconstructed higher-resolution CT images from their low-resolution counterparts with state-of-the-art performance. However, they did not investigate the possibility of using similar generative models to improve radiomics reproducibility in low dose CT.

In the topic of medical image denoising, many effective alternative procedures have been proposed to improve low dose image quality and recover texture information, including building a more sophisticated imaging platform [15], denoising in the CT sinogram domain [16], and denoising in the reconstructed CT image domain. The most convenient and popular denoising methods for low dose CT images operate in the CT image domain because hardware details and sinograms of CTs are hard to access for most researchers [23]. Generally speaking, since higher image quality should lead to higher reproducibility of radiomics, denoising seems to be a useful pre-processing step to consider.

Many articles describing image denoising techniques have been published so far, as shown in reviews [18][19]. They can be divided into two main categories: traditional denoising methods and deep learning-based denoising. The latter views denoising as a type of restricted image-to-image conversion task. Traditional denoising methods [20] have known limitations such as loss of detail in images. With the advent of deep learning, a series of publications have shown that generative models outperform traditional methods in low dose CT denoising [21][22-24]. The most widely-tested generative models are autocoders (AE) [25][26], encoder-decoder networks (EDN) [21][22], fully convolutional neural networks (FCN) [24][27] and various GANs [23][28].

An AE is an unsupervised neural network that learns how to efficiently compress (encode) an image by learning how to reconstruct the image from a tightly compressed (encoded) representation. EDNs are thus a convolutional version of AEs that allow connections across encoder and decoder layers. FCNs replace the fully connected layer in traditional convolutional neural networks with a deconvolution layer to perform the image-to-image translation.

GANs were originally proposed by Goodfellow *et al.* [28], and have been applied to diverse image-to-image translation tasks [29][30]. One of the major disadvantages of original GANs was the difficulty to control the output. However, Mirza *et al.* [31] proposed conditional GANs (CGANs), which introduced conditional restrictions into GANs to make the output more controllable and training more stable. Yang *et al.* [23] then achieved state-of-art performance in low dose CT denoising using a CGAN with Wasserstein distance penalty and perceptual loss. The general limitation of deep learning based methods remains high computational resource requirements during training and the need for large datasets.

In this article, we focus on using generative models to improve the image quality of low dose CT images and assess the impact it has on the reproducibility of radiomic features. Source code of the whole project with detailed instructions, pre-trained models and supplementary materials are available (https://gitlab.com/UM-CDS/low-dose-ct-denoising/-/branches). We hope our codes and related documents could assist future researchers to interpret and re-use our work.

## 2. Methods and Materials

Institutional Review Board approval was not applicable for this study, since the primary source of data was an open access collection on The Cancer Imaging Archive (National Institutes of Health) [32] and all patients' private information had been moved from CT scans. This dataset has been used for this study in accordance with the Creative Commons Attribution-NonCommercial 3.0 Unported (CC BY-NC) conditions.

### 2.1 Model Development

Though FCNs have shown good performance in image-to-image translation tasks, we excluded them from our experiments because they are unfeasibly slow, since they generate new images pixel-by-pixel. As convolutional versions of AEs, EDNs are expected to have better performance for image-to-image translation tasks [22]. Therefore, we excluded AEs from our experiments in favour of EDNs. CGANs were included in the experiments due to proven performance for low dose CT denoising work [23].

The architecture of our EDN is shown in Figure 1. It is a 5-layer network consisting of 32 (3×3)-sized convolution and deconvolution kernels with padding to keep the image size constant after each convolution or deconvolution. Max-pooling layers are used with 2×2 size filters and a stride of 2. We used cross entropy as the loss function and leaky rectified linear units (LReLUs) as activation functions. An original 512×512×1 CT image was fixed as the constant dimension of input and output images when training and testing.

For our CGAN, we used the same architectures and parameter settings as proposed elsewhere [31][33]. The architecture of the CGAN is illustrated in Figure 2. We adjusted the network's input and output dimensions from the original to 512×512×1. Finally, we adjusted the networks to output DICOM files directly for archival and radiomics feature calculation. The loss function in the CGAN was also set to cross entropy.

In order to compare the performance of generative models with that of traditional denoising methods, we included a type of low-pass filter, non-local means algorithm [34], as a good representative for traditional denoising methods. There are a couple of reasons to choose non-local means as our comparison denoising method. First, non-local means had the better performance amongst other traditional denoising methods [35]. Second, non-local means executed faster than other algorithms such as Block-Matching and 3D filtering (BM3D) [35][36] but gave similarly denoising outcome. For our non-local means algorithm, we set the size of the 'search windows' to 5 and the filtering parameter 'h' to 7.

### 2.2 Data Acquisition

We used the high quality NSCLC-Radiomics collection [37] (hereafter called LUNG 1), which contains CT scans of 422 non-small cell lung cancer (NSCLC) patients, as our experimental dataset. These CT scans included annotations drawn by specialist radiation oncologists that delineate a region of interest (ROI), the gross tumor volume. ROIs were necessary to be able to compute radiomic features. The CT images for which the dose level ('parameter exposure' in DICOM metadata) was missing (n=200) were excluded from further analyses. We considered CT images scanned at 400 milliampere-seconds (mAs) and above as full dose CT (n=157, the index of LUNG 1 patients included in the experiments can be found in Supplementary Table 1, supplementary materials are available: https://gitlab.com/UM-CDS/low-dose-ct-denoising/-/branches). These data were be used for training (n=40, 4260 frames) and testing (n=117, 13423 frames). Conversely, we designated CT images scanned at 50 mAs as low dose CT, taking the same definition as a prior Low Dose CT Grand Challenge [22][38].

As mentioned, training of EDNs and CGANs require paired images, in our case, pairs of matching low dose and full dose CT scans. However, LUNG 1 contains no paired images, thus we simulated the noisy degradation present in low dose CT images by introducing noise using the method proposed in literature [22][38]. In these, the authors had mimicked CT scanners' behavior by adding noise with a normal distribution into a sinogram (by Radon transform) and reconstructed the CT image from the modified sinogram to obtain simulated noisy images. We used a similar method to add noise in the original sinogram as follows:

$$z_i = (1 + b_i)e^i + r_i, i = 1, \dots, I, b_i \sim N(\mu, \sigma) \tag{1}$$

where $z_i$ is the measurement along the $i$-th ray path; $r_i$ is the read-out error; $e^i$ represents the original line integral of attenuation coefficients along the $i$-th ray path; and, $b_i$ is the black scanner factor, which follows a normal distribution. The intensity of noise added to the image can be controlled through the parameter $b_i$.

To simulate low dose CT images (scanned with 50mAs) from full dose CT images (scanned with 400 mAs), we first measured the noise intensity introduced in images with lower doses by scanning a Gammex 467 CT phantom (Middleton, WI, USA) using a Philips Brilliance Big Bore CT at two dose

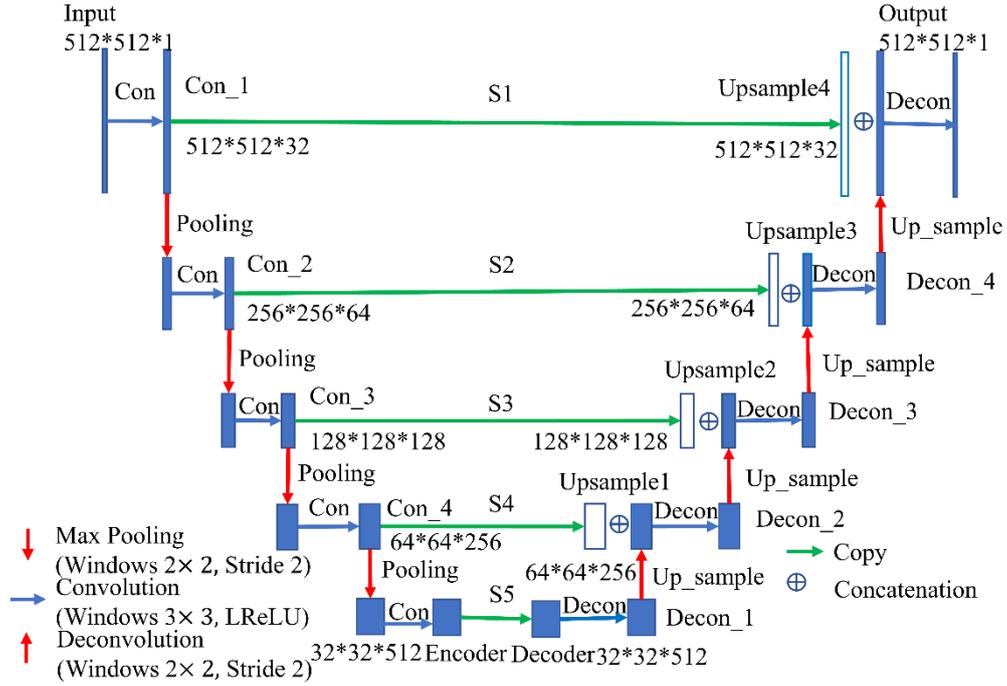

Figure 1.The architecture of the encoder-decoder network

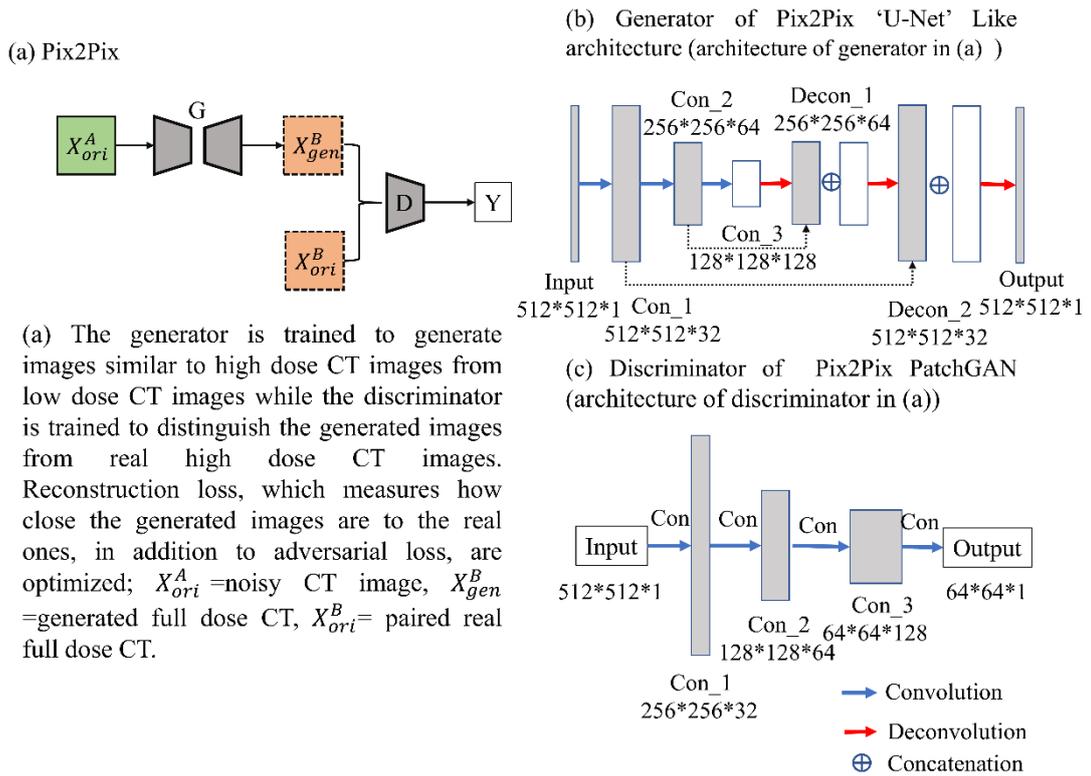

(a) Pix2Pix

(b) Generator of Pix2Pix 'U-Net' Like architecture (architecture of generator in (a) )

(a) The generator is trained to generate images similar to high dose CT images from low dose CT images while the discriminator is trained to distinguish the generated images from real high dose CT images. Reconstruction loss, which measures how close the generated images are to the real ones, in addition to adversarial loss, are optimized; $X_{ori}^{A}$ =noisy CT image, $X_{gen}^{B}$ =generated full dose CT, $X_{ori}^{B}$ = paired real full dose CT.

(c) Discriminator of Pix2Pix PatchGAN (architecture of discriminator in (a))

Figure 2.The architecture of the CGAN (Pix2Pix)

levels (400 mAs and 50 mAs) [39]. The signal-to-noise ratio (SNR) of the real phantom dataset was 19.7 dB (95%CI [17.8, 21.6]). We thus estimated that a σ value of 0.0035 best estimated the noise in 50 mAs CT images when generated from 400 mAs images. The SNR in the simulated low-noise images was 18.3 (95%CI, [16.9, 20.1]) dB, close to the real value. To assess the reproducibility of radiomic features with noise of different intensities, we added stronger noise (25 times noise power) by setting σ to 0.0068 to mimic CT images with stronger noise (referred to as simulated high-noise images hereafter). The SNR in the simulated high-noise images had thus reduced to 6.0 (95%CI, [5.9,6.1]) dB. Additionally, extraneous noise introduced by the Radon transform and inverse Radon transform was filtered from the simulated images. A comparison of noise in simulated images and in real phantom scans is shown in Figure 3, the intensity of noise in real phantom is 17.1 dB and average noise power spectra density within whole image is 45.8 W/Hz. The intensity of noise in simulated low-noise images is 19.4 dB and average noise power spectra density within whole image is 3.6 W/Hz, intensity of noise in simulated high-noise images is 6.1 dB and average noise power spectra density within whole image is 6.0 W/Hz.

We used 40 subjects from LUNG 1 (4260 images in total) for training in all three denoising models and then 117 subjects from LUNG 1 for testing. Training was only based on paired low dose (high-noise) and full dose images for all models. Denoising for the low-noise images was performed using the trained models without any additional retraining.

To test the performance of our models in denoising real low dose CTs, we used two additional datasets. First, a collection of phantoms CTs were scanned at different exposure levels, as in [39]. Second, we used the Reference Image Database to Evaluate Therapy Response (RIDER) collection, a collection of same-day repeat CT scans collected to assess the variability of tumor measurements [40]. This dataset comprised paired CT scans for 32 NSCLC patients with corresponding gross tumour volume annotations. These CT images had been scanned at low doses (7 to 13 mAs), making it suitable tests for our denoising experiment.

### 2.3 Calculation of Radiomic Features

Radiomics features from images in DICOM format were extracted using the open source O-RAW extension [41] to PyRadiomics [42]. Radiomic features can be divided into three classes – shape features, intensity histogram (first-order) features and textural (Haralick) features. In our experiments, the ROI masks for calculating radiomic features were not affected by denoising, therefore shape features were excluded from further analysis. Finally, 90 radiomic features (listed in Supplementary Table 2, supplementary materials are available at https://gitlab.com/UM-CDS/low-dose-ct-denoising) were included for our analyses. We followed the recommendations of the Image Biomarker Standardization Initiative (IBSI). The IBSI checklist can be found in Supplementary Table 3.

### 2.4 Experiments

Experiments were executed in a virtual Amazon Elastic Compute instance (G3 Graphics Accelerated Instance with Tesla M60 GPU, 30.5GB of memory and 4 CPUs).

We executed three kinds of comparisons of radiomics feature reproducibility, comparing : (i) denoising of CT images using different types of generative models (EDN, CGAN) and one traditional denoising algorithm (non-local means), (ii) performance for different numbers of training epochs (25, 50, 75 and 100) and (iii) performance for different noise intensities (low and high noise images). We compared feature reproducibility by calculating the correlation of each radiomic feature between each full dose CT image with its a corresponding noise-added then post-denoised CT.

To assess the impact of denoising on real low dose CT scans using the models trained above on LUNG 1, we ran two additional experiments. First, we used the trained models to denoise real low dose CT scans of phantoms scanned at 50 mAs and then compared the difference between low and full dose CT to the difference between denoised and full dose CT. Second, we assessed the impact of denoising on the test-retest reproducibility of radiomic features using the aforementioned RIDER dataset. Finally, we compared the test-retest reliability of radiomic features in original versus denoised CT scans by calculating the correlation between the CT scan pairs for each radiomic feature.

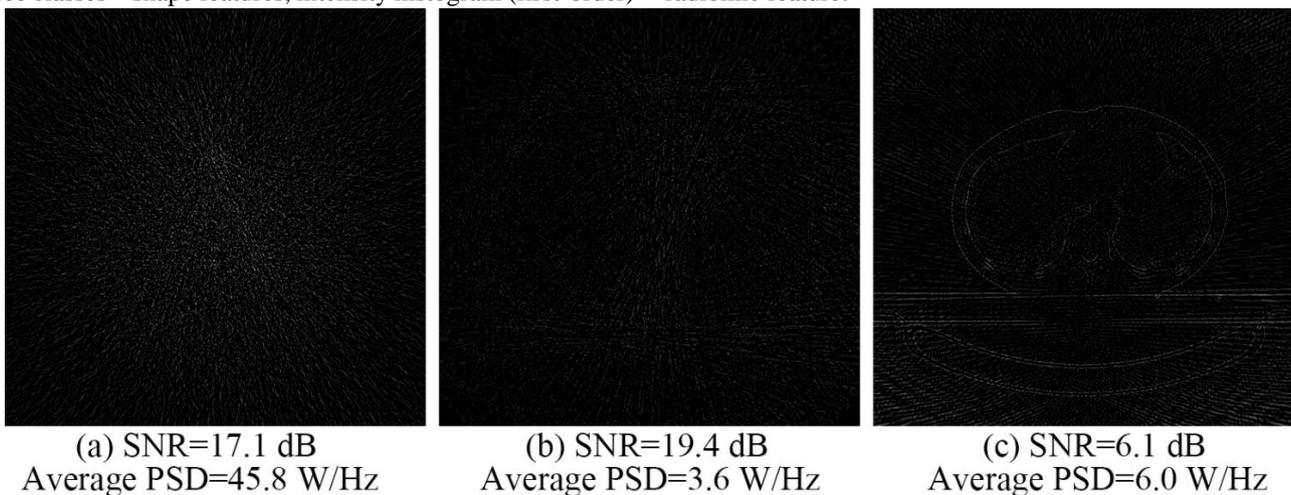

(a) SNR=17.1 dB
Average PSD=45.8 W/Hz

(b) SNR=19.4 dB
Average PSD=3.6 W/Hz

(c) SNR=6.1 dB
Average PSD=6.0 W/Hz

Figure 3. A comparison of noise in simulated images and in a real phantom image. (a) real phantom image; (b) simulated low-noise image; (c) simulated high-noise image.

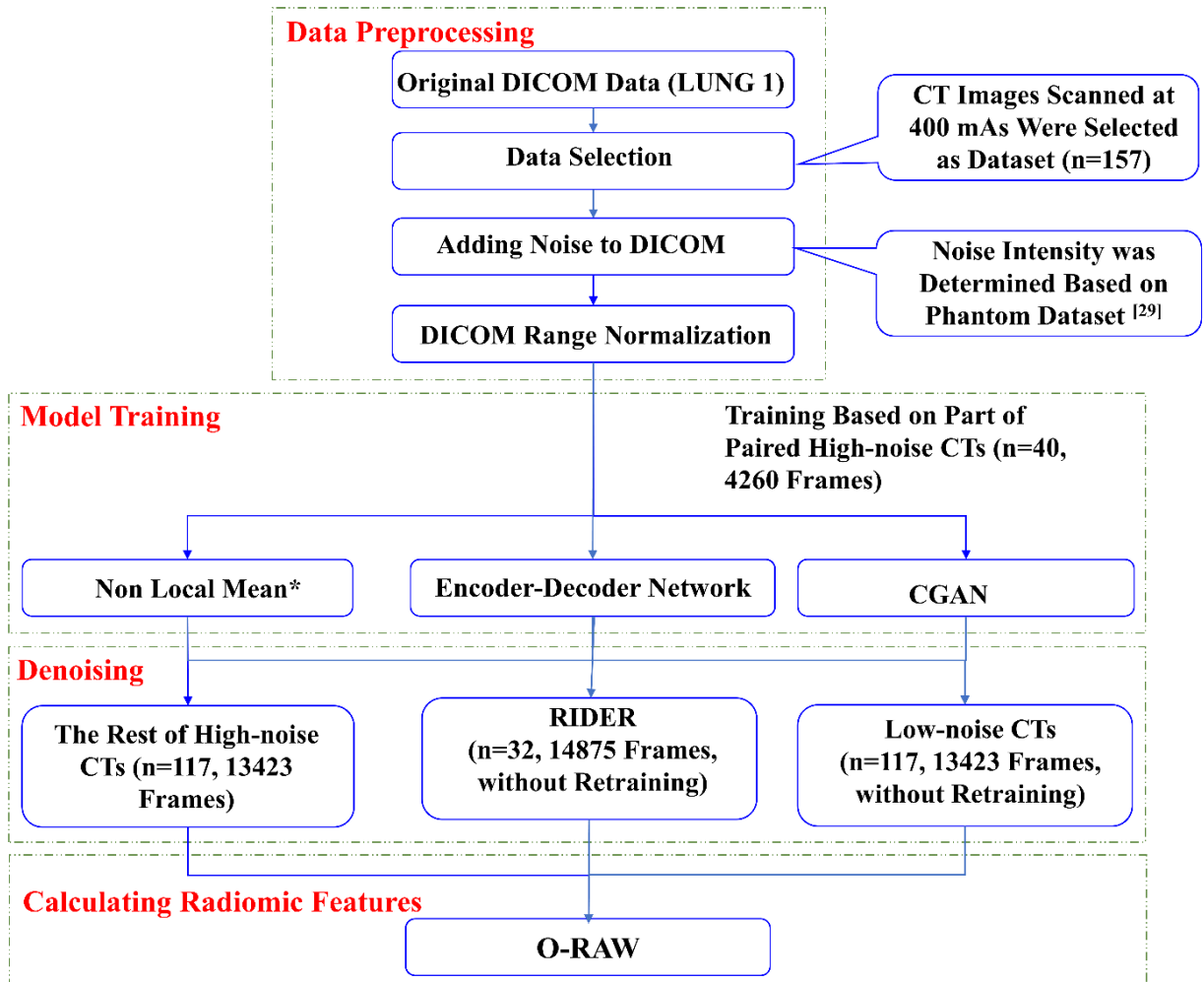

Figure 4. Flowchart of methods. *Non-local means algorithm applied only on The Rest of High-noise CTs and Low-noise CTs datasets.

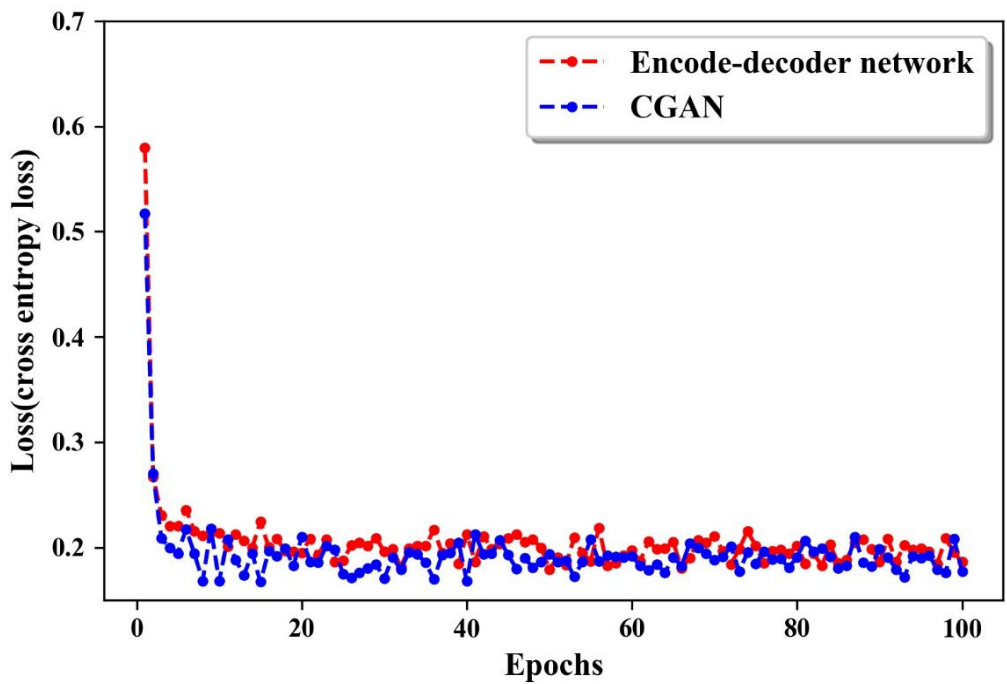

Figure 5. Loss function (cross entropy loss) curves of encoder-decoder network and CGAN along with different training epochs

## *2.5 Statistical Analysis*

Correlation was defined as the concordance correlation coefficient (CCC) [43]. We measured the difference between the original full dose CT images and denoised images using Root Mean Square Error (RMSE) and content loss [44]. Content loss was calculated based on a pretrained VGG-16 [45]. The definition of RMSE is shown in equation (2):

$$RMSE = \sqrt{\frac{1}{M}\sum_{i=1}^{M}(y_i - \hat{y}_i)^2} \qquad (2)$$

where $y_i$ and $\hat{y}_i$ represent the image value in position $i$ for original full dose CT and denoised CT, respectively. Image intensities $y$ and $\hat{y}$ were normalized to 0-1 before calculating RMSE. $M$ represents the total number of pixels in an image and it is 262144 (512 x 512) in our case.

The flowchart summarizing our study methodology is shown in Figure 4.

## 3. Results

Training took 8 hours for the EDN model and 20 hours for the CGAN model. The loss function (cross entropy loss) curves of EDN and CGAN as a function of training epochs is shown in Figure 5, showing similar convergence. The trained models were used to generate denoised CT scans of the test set. An example of an original, low-noise and post-denoising CT scan is shown in Figure 6. A corresponding figure for high-noise level images is given in Supplementary Figure 1.

Following the classification in [8], the reproducibility of a feature was deemed good, medium or poor when CCC≤0.85, 0.65≤CCC<0.85 and CCC<0.65, respectively. Table 1 shows the RMSE, content loss and ratio of poor-, medium-, and highly-reproducibility radiomic features. We summarized the reproducibility of radiomic features in low-noise images and their denoised counterparts using a heatmap in Figure 7 (a corresponding figure for high-noise images is shown in Supplementary Figure 2, the CCCs for each feature in different models can be found in Supplementary Tables 4-5).

## *3.1 Effect of Different Models*

As shown in Table 1, the baseline RMSE and content loss of high-noise and low-noise images (prior to denoising) were 0.0237(RMSE)/0.0781(content loss)  and  0.0225/0.0706, respectively. The RMSE and content loss decreased to 0.0175/0.0443 for the high-noise images and 0.0173/0.0427 for low-noise images, respectively, by using EDN denoising.

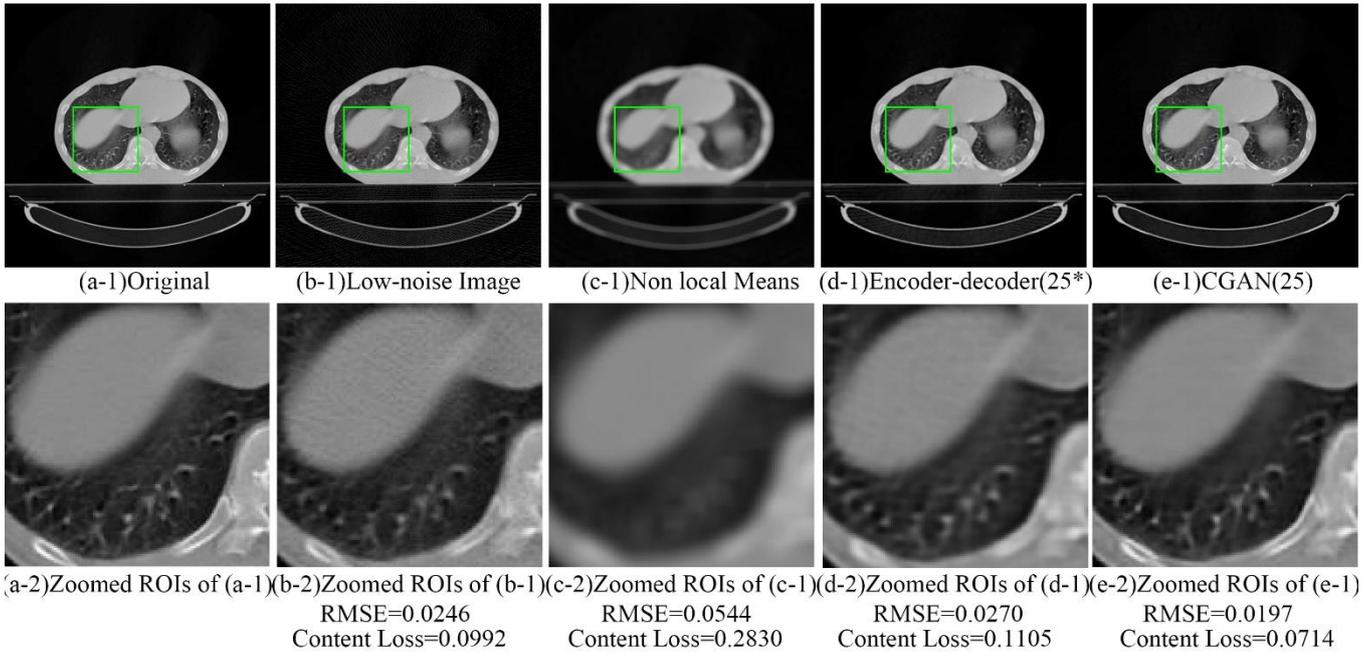

(a-1)Original    (b-1)Low-noise Image    (c-1)Non local Means    (d-1)Encoder-decoder(25*)    (e-1)CGAN(25)

(a-2)Zoomed ROIs of (a-1)   (b-2)Zoomed ROIs of (b-1)   (c-2)Zoomed ROIs of (c-1)   (d-2)Zoomed ROIs of (d-1)   (e-2)Zoomed ROIs of (e-1)

                 RMSE=0.0246         RMSE=0.0544         RMSE=0.0270         RMSE=0.0197

              Content Loss=0.0992    Content Loss=0.2830    Content Loss=0.1105    Content Loss=0.0714

Figure 6. Example of low dose CT denoising. (a-1) The original full dose CT image; (b-1) Low-noise image; (c-1) Image denoised using non-local means; (d-1) Image denoised by encoder-decoder network (Training at 25 epochs); (e-1) Image denoised by CGAN; (a-2) to (e-2) Zoomed ROIs for (a-1) to (e-1). We regard the higher noise in (d-2) by comparing with (b-2) as a coincidence.

Table 1. Summary of RMSE, content loss and distribution of CCCs of radiomic features

| Models | Distribution | RMSE | Content loss | CCCs<0.65 | 0.65≤CCCs<0.85 | CCCs≥0.85 |
|---|---|---|---|---|---|---|
| **Low-noise Images** | | | | | | |
| Without denoising | | 0.0225 | 0.0706 | 0/17(0%)* | 3/17(18%) | 14/17(82%) |
| | | | | 9/73(12%)** | 17/73(24%) | 47/73(64%) |
| | | | | 9/90(10%)*** | 20/90(22%) | 61/90(68%) |
| Non-Local Means | | 0.0993 | 0.3280 | 5/17(29%) | 8/17(47%) | 4/17(24%) |
| | | | | 48/73(66%) | 21/73(29%) | 4/73(5%) |
| | | | | 53/90(59%) | 29/90(32%) | 8/90(9%) |
| Encoder-decoder | | 0.0173 | 0.0427 | 0/17(0%) | 1/17(6%) | 16/17(94%) |
| | | | | 0/73(0%) | 16/73(22%) | 57/73(78%) |
| | | | | 0/90(0%) | 17/90(19%) | 73/90(81%) |
| CGAN | | 0.0143 | 0.0290 | 0/17(0%) | 0/17(0%) | 17/17(100%) |
| | | | | 3/73(4%) | 15/73(21%) | 55/73(75%) |
| | | | | 3/90(3%) | 15/90(17%) | 72/90(80%) |
| **High-noise Images** | | | | | | |
| Without denoising | | 0.0237 | 0.0781 | 5/17(29%) | 1/17(6%) | 11/17(65%) |

|  |  |  |  |  |  |
|---|---|---|---|---|---|
|  |  |  | 27/73(37%) | 20/73(27%) | 26/73(36%) |
|  |  |  | 32/90(36%) | 21/90(23%) | 37/90(41%) |
| Non-Local Means | 0.1095 | 0.3941 | 6/17(35%) | 7/17(41%) | 4/17(24%) |
|  |  |  | 52/73(71%) | 17/73(23%) | 4/73(6%) |
|  |  |  | 58/90(64%) | 24/90(27%) | 8/90(9%) |
| Encoder-decoder | 0.0175 | 0.0443 | 0/17(0%) | 1/17(6%) | 16/17(94%) |
|  |  |  | 4/73(5%) | 13/73(18%) | 56/73(77%) |
|  |  |  | 4/90(4%) | 14/90(16%) | 72/90(80%) |
| CGAN | 0.0146 | 0.0305 | 0/17(0%) | 1/17(6%) | 16/17(94%) |
|  |  |  | 0/73(0%) | 13/73(18%) | 60/73(82%) |
|  |  |  | 0/90(0%) | 14/90(16%) | 76/90(84%) |

\* represents the summary of fist order features; \*\* represents the summary of textural features; \*\*\* represents the summary of all features.

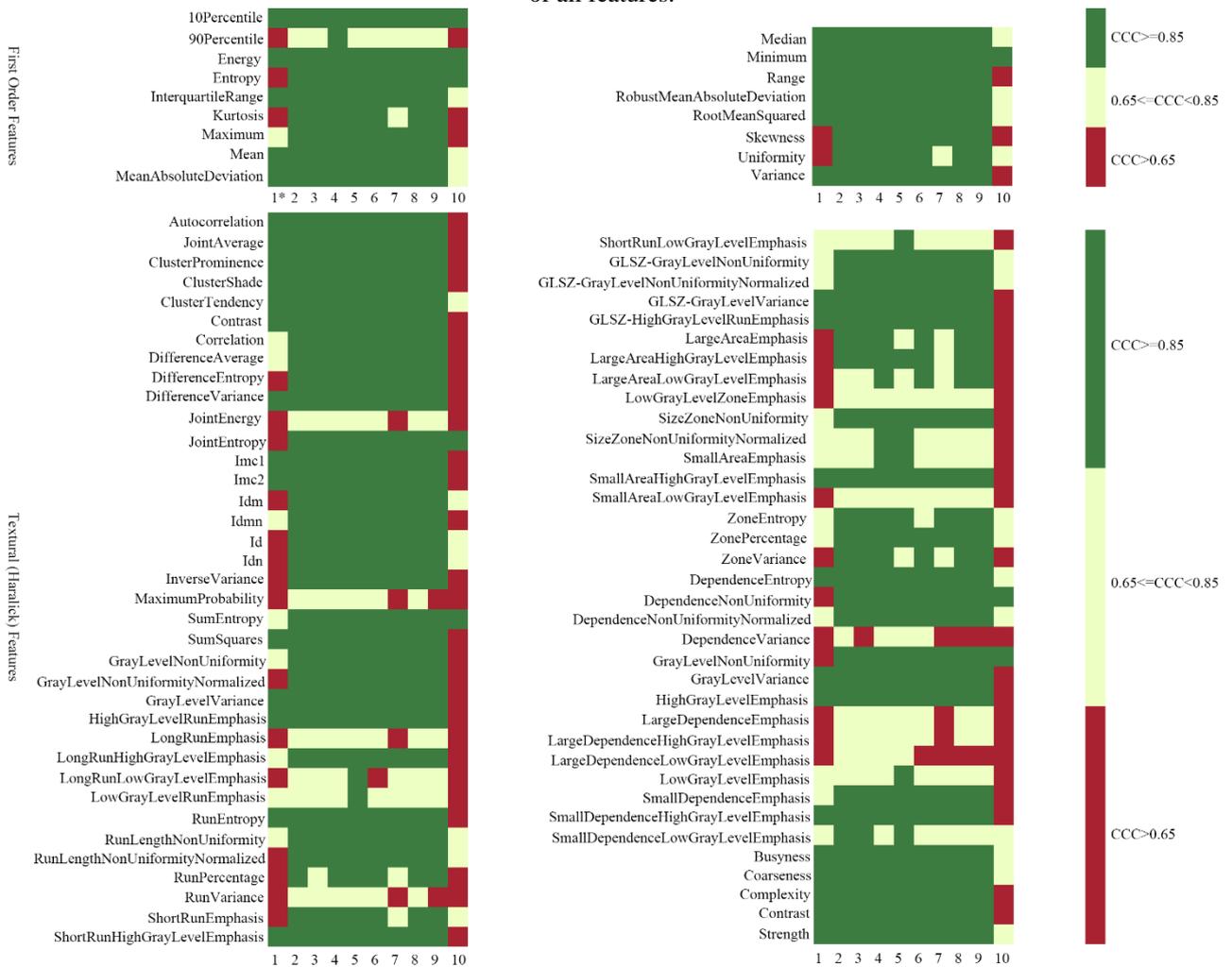

Figure 7. Heatmap of radiomic features' reproducibility based on high-noise/denoised images. \*1 represents CCC of radiomic features calculated based on high-noise images; 2-5 represent CCC of radiomic features calculated based on denoised images by using CGAN when network trained at 25, 50, 75,100 epochs; 6-9 represent CCC of radiomic features calculated based on denoised images by using encoder-decoder network when network trained for 25, 50, 75,100 epochs respectively; 10 represent CCC of radiomic features calculated based on denoised images by using non-local means algorithm.

In comparison, RMSE and content loss were decreased even further to 0.0146/0.0305 and 0.0143/0.0290 using CGAN denoising.

As shown in Figure 8, the baseline mean CCC of radiomics in high-noise and low-noise images were 0.681 [95%CI, (0.617,0.745)] and 0.867 [95%CI, (0.833,0.901)], respectively. By comparison, the mean CCC for denoised images using the EDN and the CGAN (both trained for 100 epochs) were significantly improved to about 0.92 [95%CI, (0.909,0.935)] for high-noise as well as low-noise images.

In regards to a traditional denoising method, the RMSE and content loss increased to 0.1095/0.3941 and 0.0993/0.3280 when using the non-local means algorithm. Likewise, the mean CCC of radiomics in images denoised using non-local means were decreased - 0.525 [95%CI, (0.474, 0.576)] and 0.555 [95%CI, (0.507, 0.604)] - for high-noise and low-noise images, respectively.

A cumulative distribution function of CCCs for different models when trained for 100 epochs is shown in Figure 8. The EDN and the CGAN both improved the overall reproducibility of radiomic features significantly, especially in the high-noise images. Non-local means algorithm was able to remove noise from images as we can see in Figure 6 (c-1), however it also led to detail loss, as shown in Figure 6 (c-2). This is a well-known compromise of denoising referred to as "smoothing" in literature [5]. In other words, smoothing due to a specific low-pass noise filter in the traditional method caused a deterioration of radiomics reproducibility when measured by its CCC (Wilcoxon signed-rank test, p-value <0.01).

### 3.2 Effect of Different Numbers of Training Epochs

An example of original, noisy and denoised CT scan after different training epochs by EDN and CGAN is shown in Figure 9. A cumulative distribution function of the CCCs of radiomic features on images denoised with the CGAN trained for different numbers of epochs is shown in Figure 10.

This indicated the best results in low-noise images were achieved when the CGAN was trained for 25 epochs. However, there was no significant difference in high-noise images. This same observation is holds for the EDN (Supplementary Figure 3), where the results seemed poorest when trained for 50 epochs. We speculate that this dip was an artefact of training and model convergence, rather than any meaningful finding. Table 2 shows the RMSE, content loss and proportion of poor, medium, and good reproducibility radiomic features denoised with CGAN, as a function of training epochs. (A corresponding table for the EDN is in

Supplementary Table 6). The RMSE and content loss in the whole dataset for images denoised using CGANs trained for different numbers of epochs are also shown in Table 2.

### 3.3 Effect of Different Noise Intensities

As mentioned in Section 2, the models were not retrained for denoising low-noise images. Figure 11 shows the cumulative distribution functions of CCCs of radiomic features extracted from images to which different levels of noise intensity had been applied. We compared the CCC distributions of radiomic features calculated on images denoised from high-noise images with those of images denoised from low-noise images using the Wilcoxon signed-rank test. The p-value for the CGAN and the EDN were 0.671 and 0.109, respectively, implying no significant differences. That is, our results show CGANs and EDNs trained to denoise high-noise images can also be applied to denoise images with variable levels of noise with comparable performance. A single well-trained model can be used to improve radiomics reproducibility in images with different levels of noise, and especially when the imaging dose is lower than 50 mAs.

### 3.4 Effect in real low dose CT scans

The RMSEs of denoised versus full dose CT scans of phantoms using the EDN and CGAN were 0.0182 and 0.0140 respectively, which was better than 0.0231 in the original low dose CTs. The content loss in CT scans denoised using EDN and CGAN was also improved - 0.0433 and 0.0289, respectively - compared to 0.0702 in the original low dose CTs.

The results in Table 3 show that denoising using the EDN and the CGAN improved the mean CCCs of radiomic features in the RIDER dataset from 0.89 [95%CI, (0.881, 0.914)] to around 0.94 [95%CI, (0.927,0.951)], and the percentage of features with a CCC higher than 0.85 increased from 80% to around 90%. The cumulative distribution of CCCs for radiomic features in the RIDER test set is given in Figure 12. An example of an original, denoised RIDER image using EDN and CGAN is shown in Figure 13. This scan proved especially troublesome during previous experimentation with cycle GANs (not in the scope of this paper), so it was excluded from the analysis.

We may conclude that these generative models can improve the test-retest reliability of radiomic features calculated from real low dose CT scans, such as the ones in the RIDER dataset.

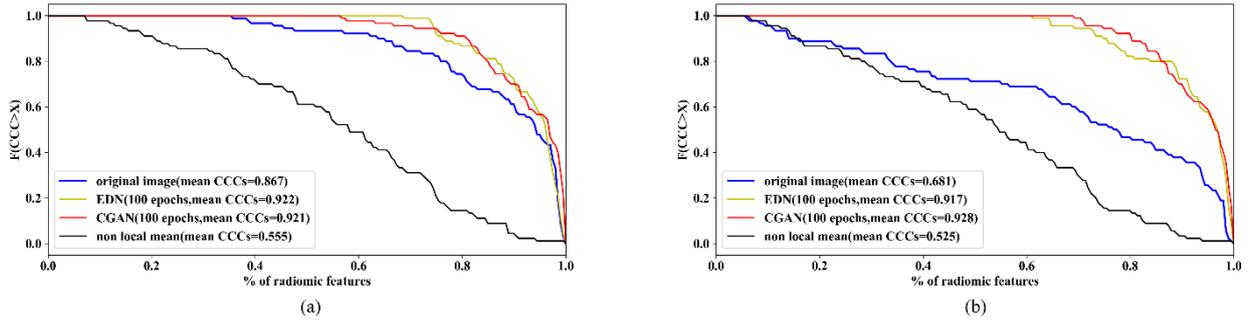

(a)                                                                                (b)

Figure 8. Cumulative distribution function of radiomic features' CCCs on images denoised using different models: (a) low-noise images; (b) high-noise images. The plots show the proportion of radiomic features with a CCC higher than x, across all possible values of the CCC (horizontal axis). For example, a point at (0.8, 0.7) implies that 70% of the radiomic features have a CCC higher than 0.8.

Table 2. RMSE, content loss and ratio of poor, medium, and good reproducibility radiomic features for images denoised by the CGAN trained for different numbers of epochs

| Training length Noisy images | 25 Epochs | 50 Epochs | 75 Epochs | 100 Epochs |
|---|---|---|---|---|
| **Low-noise Images** | | | | |
| RMSE | 0.0148 | 0.0144 | 0.0142 | 0.0143 |
| Content loss | 0.0295 | 0.0290 | 0.0276 | 0.0290 |
| CCCs ≥ 0.85 | 81/90(90%) | 73/90(81%) | 70/90(78%) | 72/90(80%) |
| 0.65≤CCCs<0.85 | 9/90(10%) | 17/90(19%) | 18/90(20%) | 15/90(17%) |
| CCCs<0.65 | 0/90(0%) | 0/90(0%) | 2/90(2%) | 3/90(3%) |
| **High-noise Images** | | | | |
| RMSE | 0.0150 | 0.0148 | 0.0144 | 0.0146 |
| Content loss | 0.0309 | 0.0312 | 0.0291 | 0.0305 |
| CCCs > 0.85 | 72/90(80%) | 71/90(79%) | 75/90(83%) | 76/90(84%) |
| 0.65≤CCCs<0.85 | 18/90(20%) | 18/90(20%) | 15/90(17%) | 14/90(16%) |
| CCCs<0.65 | 0/90(0%) | 1/90(1%) | 0/90(0%) | 0/90(0%) |

Table 3. Effect of denoising on test-retest reliability of radiomic features

| Epochs CCCs>0.85 | 25 Epochs | 50 Epochs | 75 Epochs | 100 Epochs | Original RIDER |
|---|---|---|---|---|---|
| Encoder-decoder | 78/90(0.923[*]) [(0.906,0.940)][**] | 82/90(0.936) [(0.921,0.951)] | 78/90(0.925) [(0.911,0.939)] | 78/90(0.906) [(0.881,0.930)] | 72/90(0.897) [(0.881,0.914)] |
| CGAN | 81/90(0.928) [(0.911,0.945)] | 81/90(0.920) [(0.902,0.938)] | 85/90(0.939) [(0.927,0.951)] | 83/90(0.929) [(0.914,0.944)] | |

[*]Mean CCCs of radiomic features, [**] Mean 95% confidence intervals of CCCs.

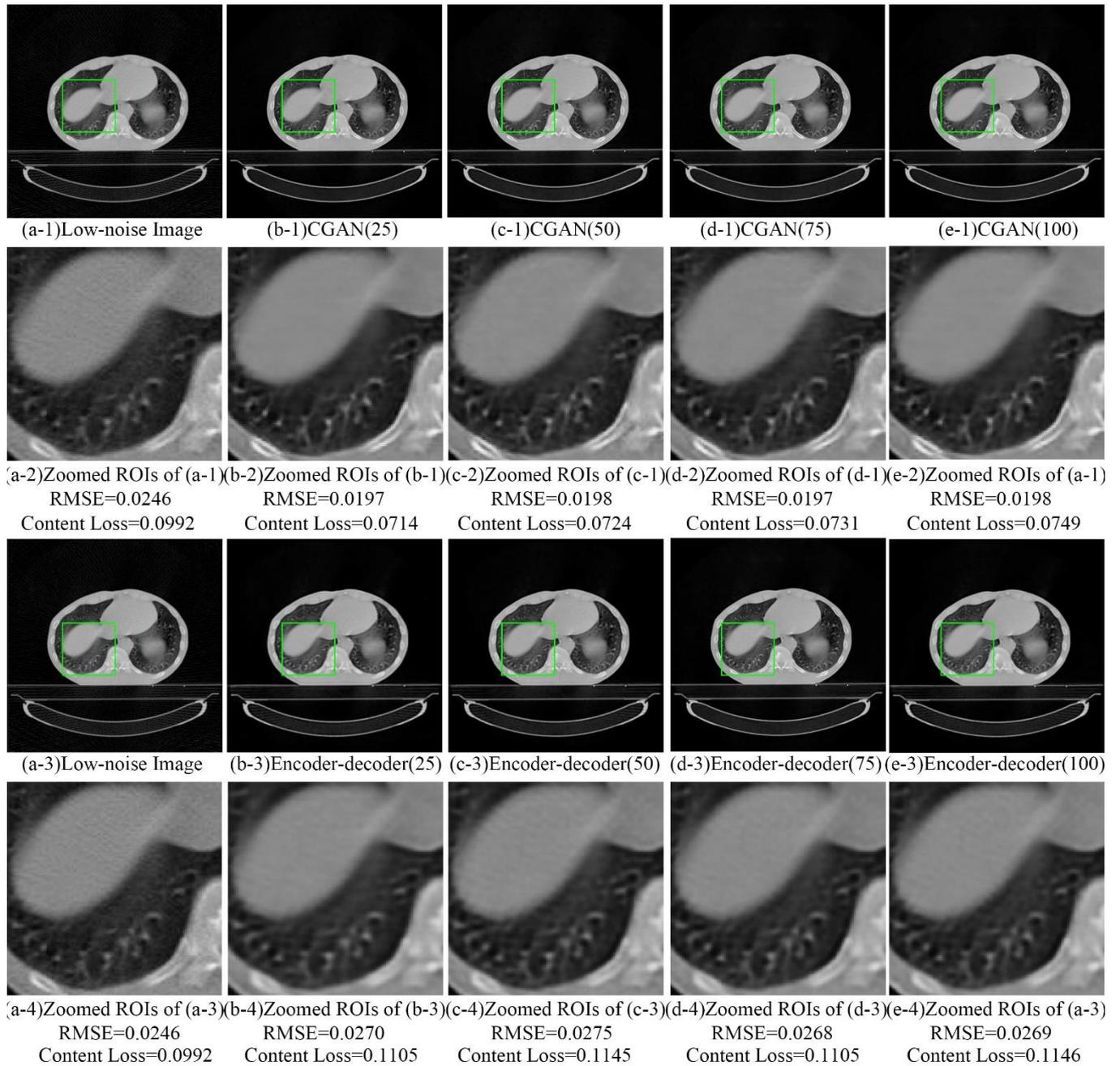

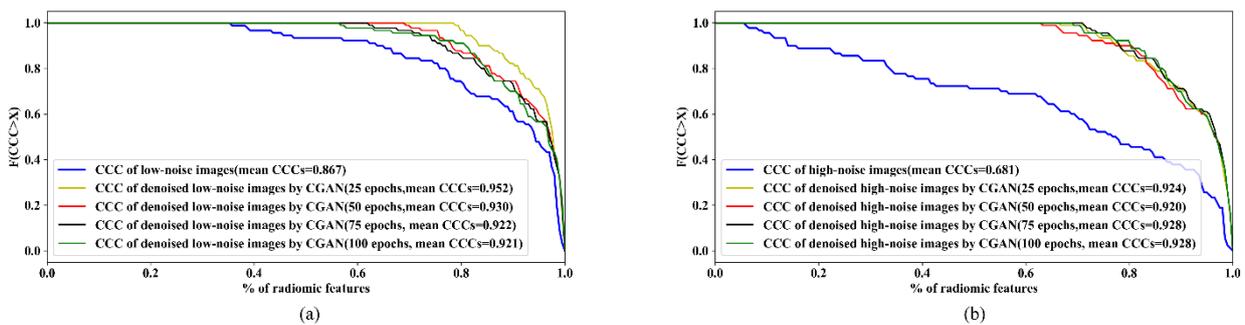

Figure 9. An example of an original, noisy and denoised CT scan at different training epochs by using the encoder-decoder network and the CGAN. As we can see figures (b-4) to (e-4), the RMSE and content loss of denoised images is higher in this particular case than the original low-noise images.

Figure 10. Cumulative distribution function of CCCs for image denoised by CGAN trained for different numbers of epochs. (a) Cumulative distribution function of CCCs based on denoised low-noise by using CGAN trained for different numbers of epochs; (b) Cumulative distribution function of CCCs based on denoised high-noise by using CGAN trained for different numbers of epochs.

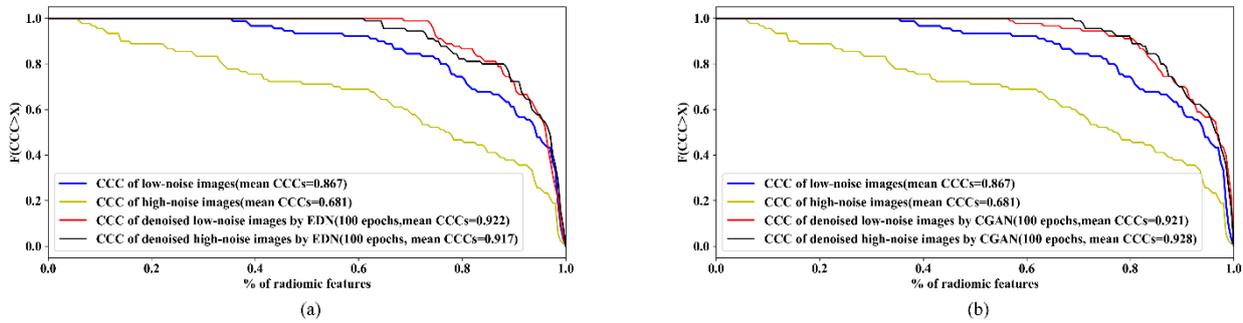

Figure 11. Cumulative distribution functions of CCCs for image noised with different intensities. (a) Cumulative distribution functions of CCCs for image noised with different intensities (Encoder-decoder network); (b) Cumulative distribution functions of CCCs for image noised with different intensities (CGAN).

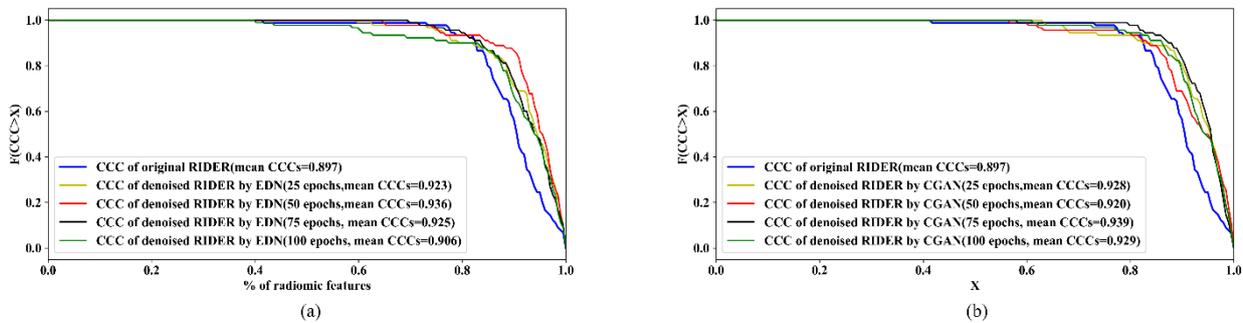

Figure 12. Cumulative distribution functions of CCCs for original and denoised CT scans in the RIDER dataset (a) using a CGAN trained for different numbers of epochs; (b) using an encoder-decoder network trained for different numbers of epochs

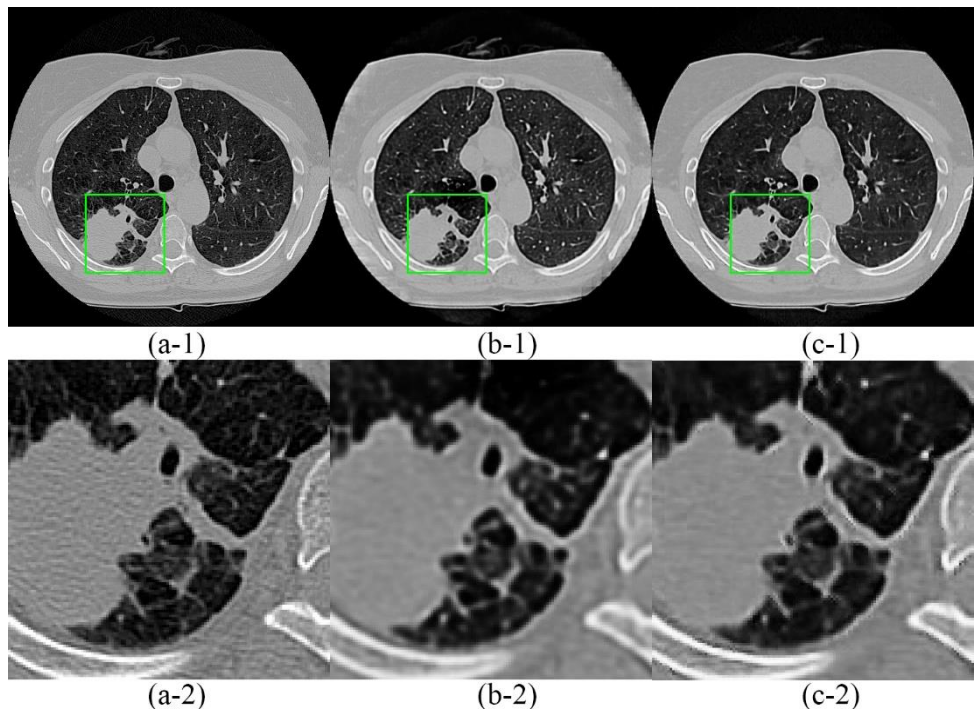

Figure 13. Example of denoised image from the RIDER dataset. (a-1) Original image; (b-1) Image denoised by EDN (100 epochs); (c-1) Image denoised by CGAN (100 epochs); (a-2) to (c-2) Zoomed ROIs for (a-1) to (c-1).

## 4. Discussion

Our objective was to test two different deep learning generative models, EDN and CGAN, to improve the SNR in CT images and explore its effect after denoising on increasing radiomic features reproducibility. The overall results of our experiments show that an equally good performance, in terms of reducing RMSE and content loss, as well as increasing the average CCC of radiomic features, was obtained by the CGAN and the EDN. However, the poor performance of the non-local means algorithm likely stems from the reduced ability of the traditional algorithm to keep fine image details during denoising, relative to CGANs and EDNs, which is the point made in Figure 6.

We chose the CCC as our metric for reproducibility rather than the intraclass correlation coefficient (ICC) [46] because it is well suited to paired before-and-after values, and it relies on fewer assumptions than the ICC [47]. As a sanity check, we also calculated ICCs for a subset of these experiments, and found that they were equal to their respective CCCs up to the second decimal place. Therefore, we considered the additional reporting of ICCs to be of limited added value.

The non-local means algorithm showed poorer relative performance against EDN and CGAN in both aspects of noise removal and content loss, as can be clearly seen by CCC of radiomics feature subgroups in Table 1 and also by visual inspection in Figure 6. The non-local means method appears to have moderate performance in maintaining first order features' reproducibility. We speculate, however, that it is the content loss (i.e. aforementioned "smoothing" phenomenon, resulting in loss of fine details from the image) that is associated with significantly worse reproducibility among the subset of textural features, when using the non-local means algorithm. The lack of reproducibility among textural features due to the non-local means algorithm is also clearly apparent when comparing column 10 in the heatmap (Figure 7) with the other columns 2-9.

Among the two generative models, CGAN showed slightly better performance than EDN in terms of removing random noise and retaining image details, but in terms of radiomics reproducibility both had similar outcomes. One of the possible reasons for this might be that radiomics features are no longer sensitive to small differences in noise or detail left behind after adequate denoising [5].

The improvement in feature reproducibility after denoising of low dose CT scans (<=50 mAs) has been demonstrated above. However, it is still worth testing if our generative models can perform equally well in a wider range of scanners and imaging conditions. The results from the low-noise images suggest that the denoising models generalize to images with different noise intensity, but greater variation in the scanning and reconstruction setting are needed to establish how generalizable this is. If so, our models might significantly reduce application barriers for clinicians and radiomics researcher.

The main limitation of this study is that training data were not actually real paired low/full dose CT images taken of the same human subjects. There is the obvious difficulty of justifying and collecting such paired images in a practical clinical setting. Overcoming these practical constraints with simulated noise, we were able to show in our experiments that denoising did have a beneficial impact on real low dose (RIDER) CT scans in terms of radiomics reproducibility. Further, we did not show a direct benefit of reproducibility for any clinical application of radiomics, such as improved performance of a prediction model. This was beyond the scope of this article, but we argue that better reproducibility of radiomic features in itself is likely to improve external validity, in general, for any potential application of radiomics.

We expended only limited time on fine-tuning and exhaustive testing of hyperparameters of the generative models. We used the same hyperparameters and training strategies for CGAN used in the original setting pix2pix [33]. The results shown in our experiments might not be the best possible results achievable with these models, especially for the CGAN, but nonetheless we demonstrated the concept of improving radiomics feature reproducibility.

The loss functions used to train the generative models might not have been the most optimal for radiomics feature reproducibility either, however we did try to improve feature reproducibility independently by minimising RMSE and content loss. Choice of loss functions can significantly affect the convergence of the network, but we did not fully investigate alternative loss functions for convergence. Other than training curves, no additional measures were implemented here to guarantee the convergence of the networks.

We transformed DICOM images to PNG images to use them as the input to the networks. The transformation will result in minimal information loss due to numerical rounding errors, but we do not believe this to have had a major detrimental effect. All of our training data originated from the same single clinic, though the RIDER test images were obtained from a different hospital. The robustness and generalizability of our models needs to be extensively tested using data from multiple centers.

Lastly, it may be possible to select other quantitative metrics to evaluate the goodness of the generative model-based approaches, however only RMSE, content loss and CCC were used for this article. This may admit the possibility of apparently worse image quality after denoising when using an EDN, as shown in Figure 9.

Future work might improve the reproducibility of radiomics features even further. For example, the mask used to calculate radiomic features was drawn by a clinician on full dose CT images. However, the masks might have been different if they had been drawn based on noisy images, such as low dose CT scans. Therefore, a noise-insensitive tumor segmentation algorithm could potentially improve low dose CT radiomic feature calculation. Moreover, we inserted Gaussian noise into the sinogram for our study, but the noise distribution in real low dose CT images might not exactly be Gaussian. This could lead to an overestimation of the performance of our models. Therefore, further studies using real data are needed.

## 5. Conclusion

In this article, we compared two different deep-learning generative models for image denoising – an EDN and a CGAN – and evaluated its utility for improving radiomics feature reproducibility (assessed by the CCC metric) in noisy images such as low dose CT scans. We compared their performance to a well-established non-deep learning based denoising method – the non-local means algorithm. We added noise at two intensities to real full dose CT images to simulate different kinds of low-dose CT images. All models were trained using high-noise images, then high and low-noise images were denoised using these models for validation, without any retraining. The results show that a non-local means algorithm for denoising may not be suitable for improving reproducibility of radiomic features. EDNs and CGANs do indeed improve the reproducibility of radiomics features in post-denoised CT, and both generative methods were about equivalent in terms of nosie removal and detail retention. In addition, the results from low-noise images were not significantly different to those of high-noise images. These results imply that images with varying levels of noise can be denoised using our trained models to potentially improve the reproducibility of radiomic features. To the authors' best knowledge, this article is the first to show that improvement in the reproducibility of radiomics features is feasible based on denoising low-dose CT images.

### Acknowledgements

JC is supported by a China Scholarship Council scholarship (201906540036) and a YERUN Research Mobility Award. The other authors acknowledge funding support from the following: STRaTegy (STW 14930), BIONIC (NWO 629.002.205), TRAIN (NWO 629.002.212), CARRIER (NWO 628.011.212), and a personal research grant by The Hanarth Foundation for LW.